\documentclass[conference,a4paper]{IEEEtran}
\IEEEoverridecommandlockouts
\usepackage{cite}
\usepackage{amsmath,amssymb,amsfonts}
\usepackage{algorithmic}
\usepackage{graphicx}
\usepackage{textcomp}
\usepackage{xcolor}
\usepackage{upgreek}
\usepackage[colorlinks=true, urlcolor=blue, citecolor=blue, linkcolor=darkgray]{hyperref}

\def\BibTeX{{\rm B\kern-.05em{\sc i\kern-.025em b}\kern-.08em
    T\kern-.1667em\lower.7ex\hbox{E}\kern-.125emX}}
\newcommand\blfootnote[1]{%
  \begingroup
  \renewcommand\thefootnote{}\footnote{#1}%
  \addtocounter{footnote}{-1}%
  \endgroup
}

\begin{document}
\title{The Tilting Mode: A New Degree of Freedom for Magneto-Mechanical Resonator Sensors}
\author{
\IEEEauthorblockN{Nora Timm\IEEEauthorrefmark{1}, Florian Hartwig\IEEEauthorrefmark{1}, Jonas Faltinath\IEEEauthorrefmark{2}\IEEEauthorrefmark{3}, Sarah Reiss\IEEEauthorrefmark{2}\IEEEauthorrefmark{3}, Pascal Stagge\IEEEauthorrefmark{1} and Tobias Knopp\IEEEauthorrefmark{1}\IEEEauthorrefmark{2}\IEEEauthorrefmark{3}\\
E-mail: nora.timm@imte.fraunhofer.de}
\IEEEauthorblockA{\IEEEauthorrefmark{1}Fraunhofer Research Institution for Individualized Medical Technology and Engineering IMTE, Lübeck, Germany\\
\IEEEauthorrefmark{2}Institute for Biomedical Imaging, Hamburg University of Technology, Hamburg, Germany\\
\IEEEauthorrefmark{3}Section for Biomedical Imaging, University Medical Center Hamburg-Eppendorf, Hamburg, Germany}
}
\maketitle
\blfootnote{\noindent This work was supported by the EU (EFRE) and the State of Schleswig-Holstein, Germany through the project IMTE 2, project number: 125 24 009.} 
\begin{abstract}
Magneto-mechanical resonators (MMRs) are an emerging class of passive, wireless sensors. Their torsional oscillation mode has recently been established for sensing and tracking applications. In this work, we report the identification and characterization of a second mechanical mode, the tilting mode, that provides sensitivity along an axis inaccessible to the torsional mode, opening up a new degree of freedom for tracking and sensing with a single MMR sensor. We derive an analytical model predicting the tilting frequency as a function of the geometric and magnetic parameters of the resonator, compare the tilting mode frequency to that of the torsional mode, and obtain a characteristic frequency ratio between the torsional and the tilting mode in the small angle approximation. Experimental characterization using three-axis excitation and detection confirms the mode's existence and its directional selectivity. Notably, the three-axis frequency response shows no observable cross-coupling between the torsional and the tilting mode. We further show that the tilting mode frequency follows the predicted dependence on magnet distance, confirming the analytical model and the mode's applicability for sensing, analogous to that of the torsional mode.
\end{abstract}


\section{Introduction}
Passive, wireless sensing is of growing interest in challenging environments such as biomedical sensing and in vivo tracking, where batteries, wiring, or electronics are impractical (e.g., catheter tracking, implant monitoring). Magneto-mechanical resonators (MMRs) address this need: a permanently magnetized rotor is suspended by a thin filament in the magnetic field of a stator magnet, forming a high quality factor (Q) mechanical oscillator that can be excited and read out wirelessly through alternating magnetic fields. The characteristic natural oscillation frequency encodes information about the local magnetic field strength at the position of the rotor, enabling both sensing and spatial tracking with one sensor \cite{Gleich2023}. Related approaches such as LC resonators \cite{Huang2016,Cui2019,Masud2023} and small-scale magnetic oscillators for localization (SMOL) \cite{Fischer2024,Fischer2024b} rely on electrical or mechanical restoring forces, respectively. In contrast, MMRs exploit the magnetic restoring torque of the stator magnet, yielding higher Q-factors at small sensor dimensions \cite{Gleich2023}.

To date, all prior MMR-based work has relied exclusively on the torsional oscillation mode, a rotation of the rotor about the filament axis. This mode can be excited by magnetic field components orthogonal to the magnetic moment of the rotor in the plane perpendicular to the filament and is read out by detecting the movement of the rotor's magnetic moment in the same plane. The torsional mode has been used for temperature sensing, pressure sensing, and three-dimensional localization of the resonator \cite{Gleich2023, Merbach2025}. However, along the filament direction, this mode cannot be excited or read out, which reduces the flexibility in excitation and readout geometry and can be a limitation for pure tracking or combined tracking and sensing applications.
In this work, we identify and characterize a second mechanical mode in MMRs: the tilting mode, in which the rotor oscillates in a plane containing the filament axis. 
\begin{figure}[tbp]
	\centerline{\includegraphics[width=0.7\columnwidth]{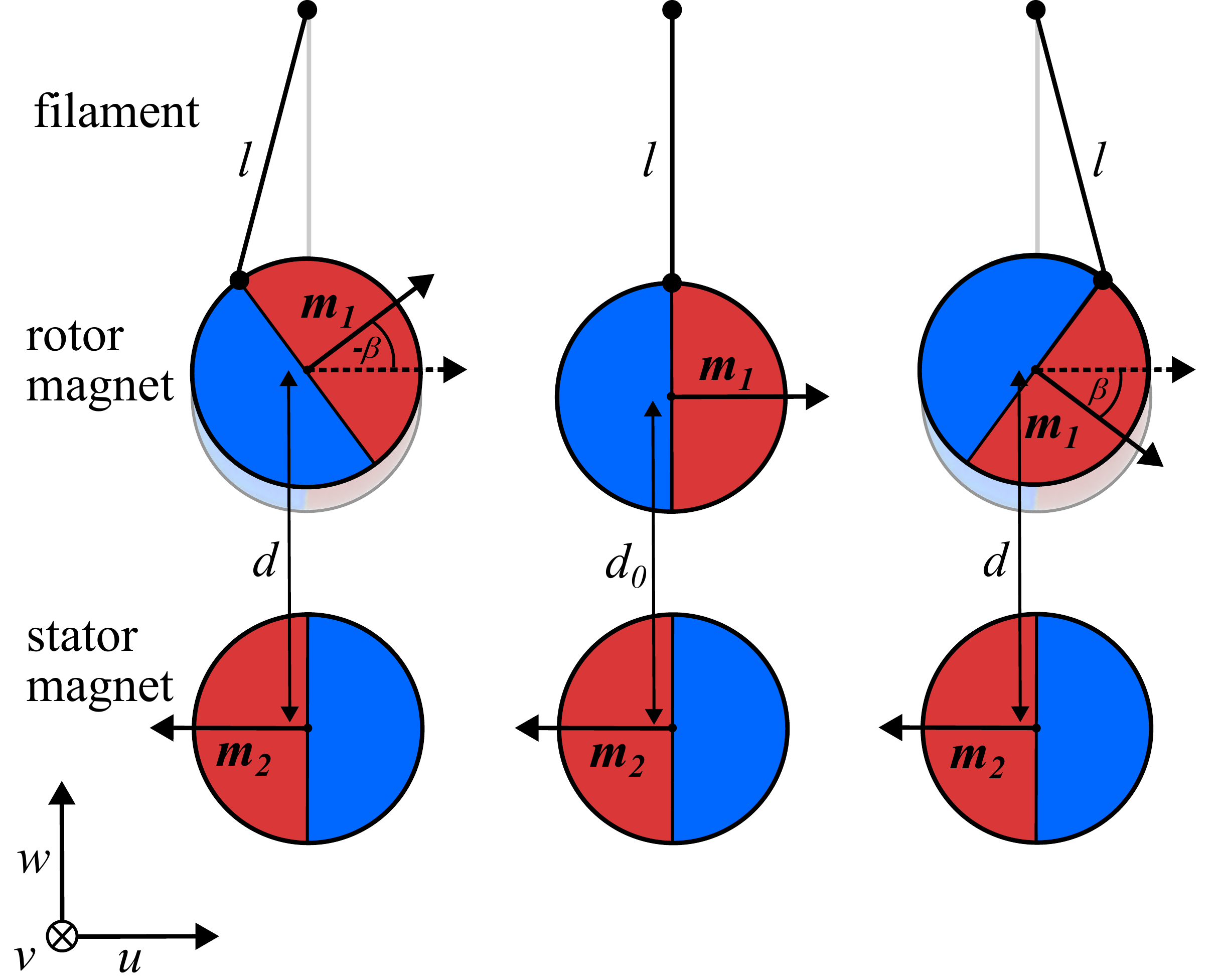}}
	\caption{Side view of the tilting mode of an MMR with filament, rotor, and stator magnet in the $uvw$-coordinate system. The central configuration is the equilibrium position with rotor and stator magnet aligned antiparallel. Left and right show deflections by an angle $\beta$ in opposite directions. The magnetic moment $m_1$ of the rotor and $m_2$ of the stator are shown, as well as the center-to-center distance $d = d(\beta)$ between both magnets, with $d_0$ denoting the equilibrium distance at $\beta = 0$. (Not to scale)}
	\label{f1}
\end{figure}
\section{Theory}\label{theo}
To identify the tilting mode kinematics, we recorded the rotor's motion using a high-speed camera (Chronos 1.4, Kron Technologies) from two different angles ($u$- and $v$-direction in Fig.~\ref{f1}) after excitation in $w$-direction. We observed that the rotor tilts about its center in the $uw$-plane while remaining centered above the stator, as shown in Fig.~\ref{f1}. Based on these observations, we formulate the following holonomic constraints for the analytical model: the rotor with radius $r$ is always centered above the stator, and the inextensible filament with length $l$ is attached at the midpoint between the magnetic poles on the surface of the rotor. Geometric analysis yields the center-to-center distance $d$ between the two magnets as a function of the tilting angle $\beta$:
\begin{equation}
	d(\beta)=d_0+r (1-\cos \beta) + l - \sqrt{ l^2 - r^2 \sin^2 \beta},
\end{equation}
where $d_0$ is the center-to-center distance in the equilibrium position. To derive the equation of motion, we model both magnets as point dipoles and express the potential energy $U$ of the rotor in the field of the stator: 
\begin{equation}
	U(\beta)=-\mathbf{m}_1\cdot \mathbf{B}_2= \frac{-\mu_0 m_1 m_2}{4\pi\; d(\beta)^3} \cos\beta,
\end{equation}
where $\mathbf{m}_1$ is the magnetic moment of the rotor, $\mathbf{B}_2$ denotes the magnetic field of the stator at the rotor position, and ${m}_1$, ${m}_2$ are the magnitudes of the respective magnetic moments. We define $K:=\frac{\mu_0 m_1 m_2}{4\pi}$ and apply the torque relation $I\ddot{\beta}=-\frac{\mathrm{d}\,U}{\mathrm{d}\,\beta}$, where $I$ is the rotor's moment of inertia, which is identical for all center-of-mass axes of a spherical rotor. After differentiation, we obtain the following equation of motion:
\begin{equation}
		\ddot{\beta}=\frac{-K}{I}\! \left(\!\frac{\sin\beta}{d(\beta)^3}\! + \!\frac{3r\sin(2\beta)}{2\; d(\beta)^4} \!\left(\!1\!+\!\frac{r \cos\beta}{\sqrt{l^2 - r^2 \sin^2 \beta}}\!\right)\!\right)\!.
\end{equation}
In the small-angle approximation $(\beta\ll 1)$, this simplifies to: 
\begin{equation}
		\ddot{\beta}\approx\frac{-K}{Id_0^3} \left(1 + \frac{3r}{d_0} \cdot \left(1+\frac{r}{l}\right)\right)\beta.
\end{equation}
The resulting expression describes a harmonic oscillation equivalent to that of the torsional mode derived in \cite{Gleich2023}, without damping, scaled by the factor $C = 1 + \frac{3r}{d_0} \cdot \left(1+\frac{r}{l}\right)$. The tilting mode natural frequency $f_{\text{tilt}}$ is thus given by
\begin{equation}
	f_{\text{tilt}}=\frac{1}{2\pi}\sqrt{\frac{K C}{ I d_0^3}}=\sqrt{C}f_{\text{torsion}},
\end{equation}
where $f_{\text{torsion}}$ is the torsional mode natural frequency \cite{Gleich2023}. The factor $C$ depends solely on the geometric ratios $r/d_0$ and $r/l$: for larger ratios, the restoring torque increases, yielding a higher tilting frequency relative to the torsional mode.

\section{Experimental Setup}\label{setup}

\subsection{Sensor Design}\label{sensor_design}

Two MMRs are used in this work. MMR$_1$ consists of two spherical NdFeB magnets (N35, $5\,$mm diameter) and an ultra-high-molecular-weight polyethylene (UHMWPE, Dyneema) filament of length $20$\,mm in a fixed housing (SLA printed, Form 4 \& clear resin V5, Formlabs). The filament is affixed to the rotor using a small cap made of the same material as the housing, bonded with cyanoacrylate adhesive. The center-to-center distance between rotor and stator is $9$\,mm.  MMR$_1$ is used for three-axis spectral characterization. MMR$_2$ consists of a spherical rotor (N40, $4$\,mm diameter), a cylindrical stator (N35, $4$\,mm diameter and $4$\,mm height), and a UHMWPE filament of length $5$\,mm. The filament is affixed to the rotor using a similar cap as in MMR$_1$. MMR$_2$ is enclosed in a compressible polyamide housing (SLS printed, Formiga P100 \& PA2200, EOS) that allows the rotor-stator distance to be varied by compressing the housing in a vise. The distance is controlled via the screw pitch of the vise, providing a known displacement of $1.5$\,mm per full revolution.

\subsection{Excitation and Readout}

The excitation and readout system is based on the transmit-receive architecture described in \cite{Mohn2025, Faltinath2025}. A cube of three orthogonal coil pairs enables independent excitation and signal acquisition in $x$-, $y$-, and $z$-direction. The MMRs are placed in the coil system such that the $xyz$-coordinate system corresponds  to the $uvw$-coordinate system of the MMRs as closely as achievable by manual placement. For each measurement, the rotor is excited with a $0.2$\,s sinusoidal burst, and the free decay signal is recorded after the excitation has stopped (MMR$_1$: for $2$\,s, MMR$_2$: for $0.3$\,s). For spectral characterization of MMR$_1$, a sequential frequency sweep from $50$ to $250$\,Hz is performed in steps of $1$\,Hz, exciting along one axis at a time with a magnetic field amplitude of $30\,\upmu$T. The free decay signal is recorded in all three axes simultaneously. The spectral response is obtained via FFT of the signal. For MMR$_2$, excitation is performed along the $y$-axis for the torsional mode and along the $z$-axis for the tilting mode, with a magnetic field amplitude of $50\,\upmu$T. The free decay signal is recorded in all three axes at various rotor-stator distances. The excitation frequency is adjusted to maximize the response. The natural frequency is obtained using the time-domain model fit described in \cite{reiss2026}.

\section{Results and Discussion}\label{results}

\subsection{Spectral Characterization}

\begin{figure}[tbp]
	\centering{\includegraphics[width=0.82\columnwidth]{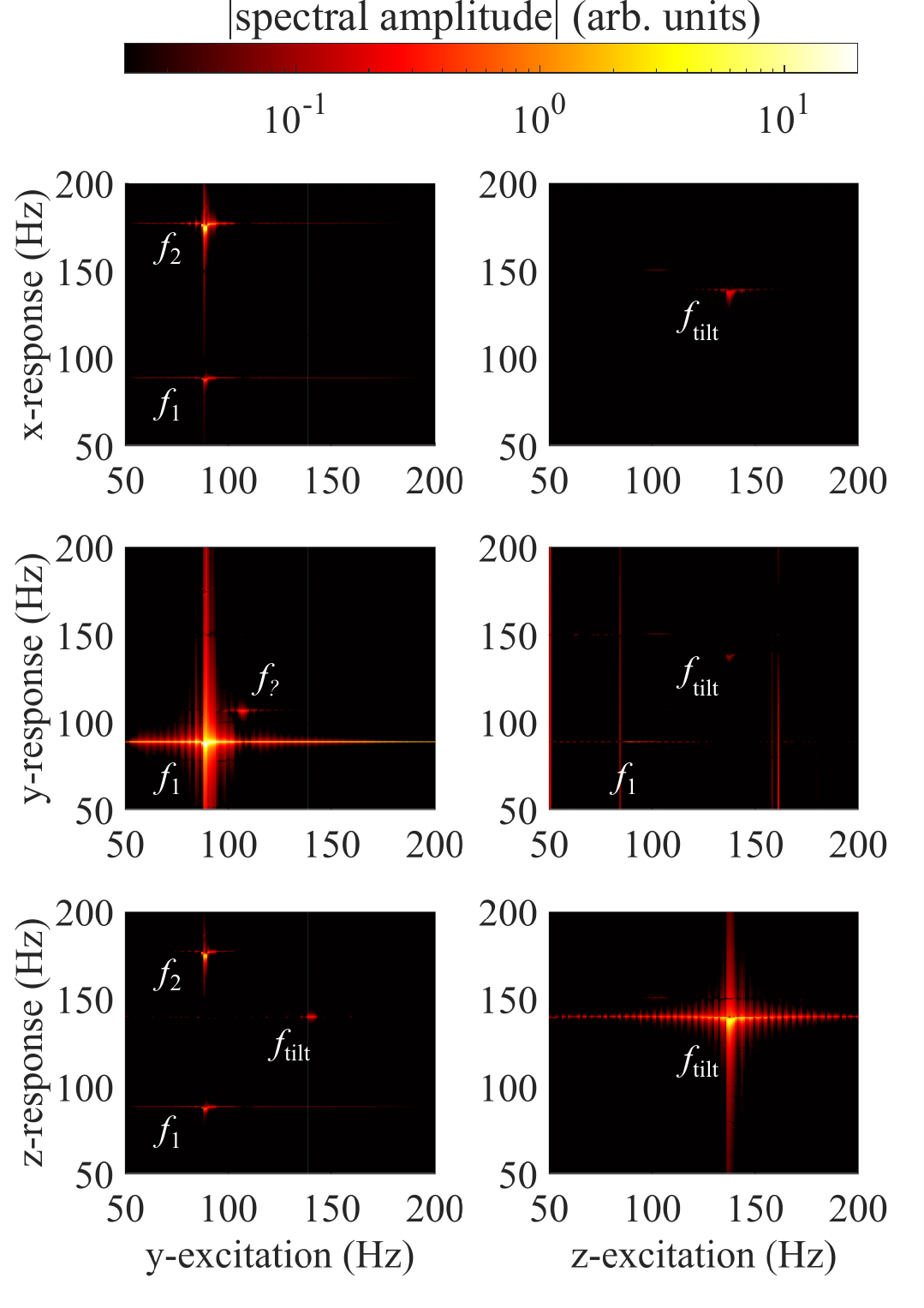}}
	\caption{Three-axis frequency response of MMR$_1$. Each panel shows the absolute value of the spectral amplitude of the free decay signal for one combination of excitation direction and response direction. The resonance peaks of the torsional mode ($f_1 = 89$\,Hz), its second harmonic ($f_2 = 177$\,Hz), and the tilting mode ($f_\text{tilt} = 138$\,Hz) are labeled, together with a weak resonance at $f_?=107$\,Hz that might be attributed to a pendulum mode. Excitation in $x$-direction is omitted as it did not yield observable resonances.}
	\label{matrix}
\end{figure}

Fig.~\ref{matrix} shows the three-axis frequency response of MMR$_1$. Each panel displays the spectral amplitude of the free decay signal for one combination of excitation and response direction. Since excitation in $x$-direction did not yield any noticeable peaks in the frequency analysis, only the results for excitation in $y$- and $z$-direction are shown. Three main resonance frequencies are clearly visible: the torsional mode at $89$\,Hz ($f_1=f_{\text{torsion}}$), the second harmonic of the torsional mode at $177$\,Hz ($f_2=2f_{\text{torsion}}$), and the tilting mode at $138$\,Hz ($f_\text{tilt}$). Besides the dominant peaks, a weak resonance close to $f_1$ is visible at $107$\,Hz ($f_?$) for $y$-excitation and $y$-response. 
The torsional mode is observed mainly for $y$-excitation and strongest in the $y$-response as expected. The second harmonic of the torsional mode is observed mainly in $x$-direction, which is in agreement with the model derived in \cite{Gleich2023}. The $z$-components of the torsional mode are attributed to slight misalignment of the MMR in the coil system and fabrication tolerances of the MMR. The tilting mode is observed mainly for $z$-excitation and strongest in the $z$-response.
However, weaker $x$- and $y$-components are also observed, as well as a slight response from $y$-excitation. Analogous to the torsional mode, these components are caused by misalignment of the MMR and fabrication tolerances. Remarkably, no cross-coupling between the torsional and tilting modes is observed: the tilting mode frequency $f_\text{tilt}$ does not appear in response to excitation at the torsional frequency $f_1$, and vice versa.
The resonance at $f_?$ might be attributed to a pendulum mode, which couples to the main torsional mode. In the recorded time signal, the additional mode is visible as a beating between the two frequencies, resulting in amplitude modulation of the signal. The oscillatory pattern of the resonance peaks is attributed to the fixed excitation duration of $0.2$\,s. Depending on the driving frequency, the phase mismatch between the driving field and the natural oscillation varies, resulting in constructive or destructive interference at the end of the excitation burst.
The ratio $f_\text{tilt}/f_1$ yields an experimental value of $\sqrt{C} = 1.55$ for MMR$_1$. Using the values for $r$, $l$ and $d_0$ from Section~\ref{sensor_design} leads to a theoretical value of $\sqrt{C} = 1.39$ for MMR$_1$. The discrepancy might be caused by neglecting the filament's bending stiffness, which provides an additional restoring torque for the tilting mode, and by the unaccounted modification of the rotor's moment of inertia due to the cap.

\subsection{Distance Series}

\begin{figure}[tbp]
	\centering{\includegraphics[width=0.9\columnwidth]{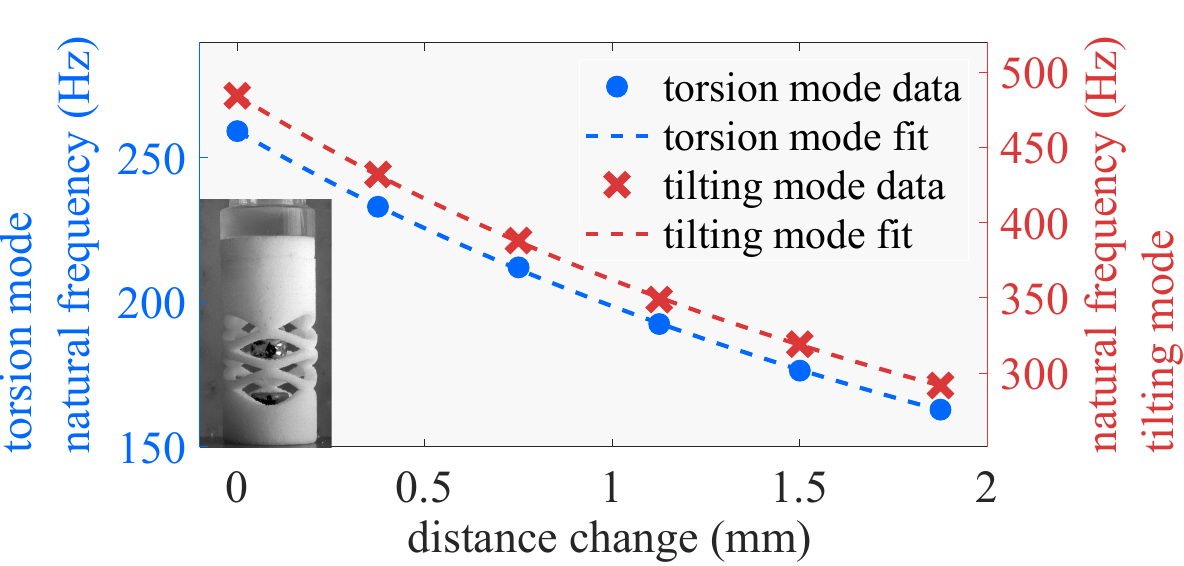}}
	\caption{Natural frequencies of the torsional mode (left axis, blue) and tilting mode (right axis, red) of MMR$_2$ as a function of rotor-stator distance change.  Dashed lines show fits of the form $a(x+x_0)^{-3/2}$ to the data. Both modes follow the same scaling with distance. The inset is a microscope image of MMR$_2$ with compressible polyamide housing.}
	\label{dist}
\end{figure}

To investigate the dependence of the tilting mode frequency
on the rotor-stator distance, the rotor-stator distance of MMR$_2$ is varied as described in Section~\ref{sensor_design}. Fig.~\ref{dist} shows the natural frequencies of both the torsional and the tilting mode as a function of the change in rotor-stator distance. Each data point represents the mean of 20 measurements. Error bars are omitted as the standard deviation is below $1.5$\,Hz for all data points. Both data sets are fitted with functions of the form $a(x+x_0)^{-3/2}$, consistent with the expected scaling \cite{Faltinath2025}. The experimental and theoretical results for $\sqrt{C}$ are shown in Tab.~\ref{tab1}. The center-to-center distance $d_0$ for each measurement is determined from the fit parameter $x_0$ of the tilting mode fit together with the screw pitch of the vise. Consistent with the results for MMR$_1$, the experimental values are systematically higher than the theoretical predictions. In addition to the factors discussed for MMR$_1$, the point-dipole approximation could be inappropriate for small values of $d_0$, since the stator of MMR$_2$ is cylindrical.

\begin{table}[tbp]
	\caption{Experimental and theoretical results for $\sqrt{C}$ as a function of center-to-center distance $d_0$ for MMR$_2$.}
	\begin{center}
		\begin{tabular}{|c|c|c|}
			\hline
            $d_0 (\text{mm})$ & $\sqrt{C}$ experiment & $\sqrt{C}$ theory \\
            \hline
             $4.65 \pm0.04$ & $1.87 \pm 0.01$ &$ 1.67 \pm 0.04$ \\
             $5.03 \pm 0.04$ & $1.85 \pm 0.01$ & $1.63\pm 0.03$\\
             $5.40 \pm 0.04$ & $1.83 \pm 0.01$ & $1.60 \pm 0.03$ \\
             $5.78 \pm 0.04$ & $1.81 \pm 0.01$ & $1.57\pm 0.03$ \\
             $6.15\pm 0.04$ &   $1.81 \pm 0.01$ & $1.54 \pm 0.03$\\
             $6.53\pm 0.04$ & $1.79 \pm 0.02$ & $1.51\pm 0.03$ \\
             \hline
		\end{tabular}
		\label{tab1}
	\end{center}
\end{table}   

\section{Conclusion and Outlook}\label{con}

We identified and characterized the tilting mode as a new degree of freedom in MMRs. The mode appears to be decoupled from the torsional mode and follows the predicted scaling with magnet distance, confirming its applicability for sensing. This opens the possibility of enhanced spatial tracking without added complexity. Future work will address the model discrepancy by incorporating the rotor's modified moment of inertia as well as the filament's bending stiffness, and investigate simultaneous dual-mode operation.

\section*{Acknowledgments}
The authors thank Dennis Wendt for designing and fabricating the compressible housing of MMR$_2$. 
\pagebreak

\bibliography{tilt_mode_ref} 

\end{document}